# Tuning the Charge Density Wave and Superconductivity in $Cu_xTaS_2$

K.E. Wagner[1], E. Morosan[2], Y. S. Hor[1], J. Tao[3], Y. Zhu[3], T. Sanders[2], T.M. McQueen[1], H.W. Zandbergen[4], A. J. Williams[1], D.V. West[1] and R.J. Cava[1]

[1]*Department of Chemistry, Princeton University, Princeton NJ 08544*
[2]*Department of Physics, Rice University, Houston TX 77255*
[3]*Condensed Matter Physics and Materials Science, Brookhaven National Laboratory, Upton NY 11973*
[4]*National Centre for HREM, Department of Nanoscience,
Delft Institute of Technology, 2628 CJ Delft, The Netherlands*

We report the characterization of layered, 2H-type $Cu_xTaS_2$, for $0 \leq x \leq 0.12$. The charge density wave (CDW), at 70 K for $TaS_2$, is destabilized with Cu doping. The sub-1K superconducting transition in undoped $2H$-$TaS_2$ jumps quickly to 2.5 K at low x, increases to 4.5 K at the optimal composition $Cu_{0.04}TaS_2$, and then decreases at higher x. The electronic contribution to the specific heat, first increasing and then decreasing as a function of Cu content, is 12 mJ mol$^{-1}$ K$^{-2}$ at $Cu_{0.04}TaS_2$. Electron diffraction studies show that the CDW remains present at the optimal superconducting composition, but with both a changed q vector and decreased coherence length. We present an electronic phase diagram for the system.

## I. INTRODUCTION

The competition between Charge Density Wave (CDW) and superconducting states at low temperatures in layered transition metal dichalcogenides has been of interest for decades. A large body of experimental and theoretical literature exists on the topic, which continues to challenge our detailed understanding of the two phenomena in this class of materials (see, *e.g.* refs. 1-13). Here we report the results of chemically tuning the CDW/superconductivity competition in tantalum disulfide, $TaS_2$, a classic, layered dichalcogenide. $TaS_2$ has several polytypes that differ in both the local coordination of the $TaS_6$ polyhedron - either octahedral or trigonal prismatic - and the number of $TaS_2$ layers in an elementary cell (14). The 1T form, with a single layer of $TaS_6$ octahedra sharing edges, is a golden semiconductor that has been made superconducting near 4.5 K by doping with Li (15). The 2H form, the subject of this study, is based on edge-sharing $TaS_6$ trigonal prisms, and is a dark gray low-temperature superconductor. The 4H, 3R and 6R polytypes are also known (14).

Undoped $2H$-$TaS_2$ is reported to have an in-plane charge density wave (CDW), q = (0.338,0,0), with a transition temperature to the CDW state of 70 K (16,17). It is also superconducting, with a $T_c$ near 0.8 K. (18,19). While there are no previous reports of the effects of Cu doping on the superconductivity of $TaS_2$, there are reports that doping with other elements yields a higher temperature superconductor. Intercalation of Na into $2H$-$TaS_2$ to form $Na_xTaS_2$ ($0 < x < 0.10$), for example, shows an increase in $T_c$ to 4.4 K (20). Further, the resistive and spectroscopic signatures of the CDW weaken on Na doping (20,21). $Fe_xTaS_2$ was found to have a $T_c$ of approximately 3.5 K when x = 0.05, with $T_c$ fully suppressed for x > 0.10 (22). There are also early reports of superconductivity in the range of 2-5 K for $2H$-$TaS_2$ doped with many organic molecules (23,24). In all these cases, the dopants are intercalated into the van der Waals layer between the $TaS_2$ planes.

None of the previous studies report a system for which $TaS_2$ has been tuned through chemical doping in small increments to follow the competition between the CDW and superconducting states. Here we report that this can be accomplished in simple equilibrium syntheses through doping with Cu. Copper is chosen because it was observed to behave as an n-type dopant in the chemically similar $Cu_xTiSe_2$ system, where it is monovalent and nonmagnetic (i.e. $3d^{10}$, S = 0) (25). We find that the superconducting transition temperature in $Cu_xTaS_2$ first increases and then decreases with Cu doping, tracing out a classical dome in $T_c$ vs. x, though the details of how this happens are unexpected. The CDW state is concurrently destabilized, though it remains present at a somewhat different wavevector and shorter coherence length at the optimal superconducting composition.

## II. EXPERIMENT

Stoichiometric amounts of elemental Cu (cleaned in dilute HCl), Ta, and S powder were mixed to yield 1 g samples of $Cu_xTaS_2$ for $0 \leq x \leq 0.14$. The powder mixtures were sealed in evacuated silica tubes. The initial heating cycle began with 550 °C overnight. Samples were air quenched to room temperature. Powders were then ground, pressed into pellets, and resealed in evacuated silica tubes. The tubes were heated at 650 °C (2 nights), 750 °C (2 nights), and then again at 650°C (2 nights), and air quenched to room temperature. A third annealing was then performed: all samples for x > 0.03 were annealed for one week at 650°C, producing single phase 2H-$TaS_2$ materials but samples for x < 0.03 were annealed at 600 °C to avoid the formation of the 3R polytype.

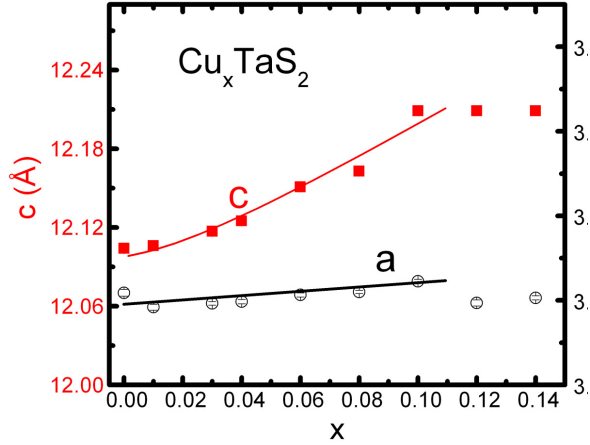

**Fig. 1** (Color online) Variation of the crystallographic cell parameters with copper content in 2H-$Cu_xTaS_2$. Standard deviations on the cell parameters are smaller than the points.

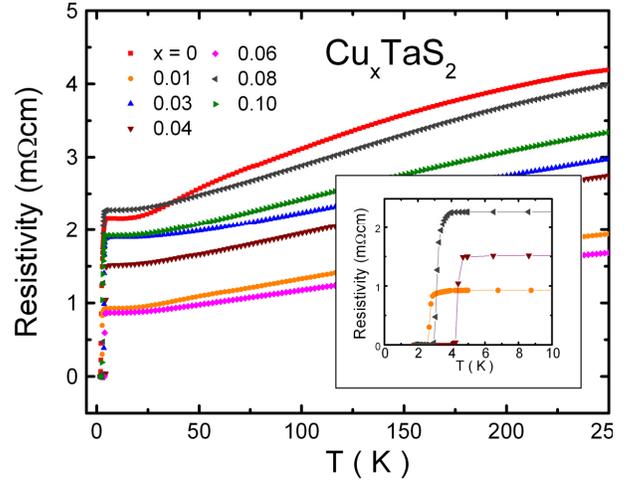

**Fig. 2** (Color online) Temperature dependent resistivity on polycrystalline pellets of $Cu_xTaS_2$. Inset, detail of superconducting transitions in $Cu_xTaS_2$ for x = 0.01, 0.04 and 0.08.

The identity and phase purity of the samples was determined by X-ray powder diffraction. Room-temperature data was recorded with a Bruker D8 diffractometer, using Cu Kα radiation and a graphite diffracted beam monochromator. The superconductivity of the $Cu_xTaS_2$ samples was characterized through temperature dependent dc magnetization measurements in a Quantum Design PPMS, which was also used to perform temperature dependent resistivity measurements using a standard four-point probe technique. Specific heat measurements were also performed in the PPMS, by using the thermal relaxation method. The Seebeck Coefficient was measured using a homemade alteration of an MMR technologies SB100 Seebeck measurement system.

Electron diffraction (ED) studies were performed on lightly ground samples mounted on copper grids coated with holey carbon film. ED experiments were carried out using the JEOL 3000F transmission electron microscope (TEM) equipped with a Gatan liquid helium cooling stage that enables *in-situ* TEM observations with the sample temperature from 11 K to 400 K. ED patterns were recorded on 16 bit Fuji imaging plates for digital analysis. Peak positions and widths on the ED patterns were quantified using standard curve fitting methods.

### III. RESULTS

The observed powder X-ray diffraction patterns for the $Cu_xTaS_2$ samples were fit to the 2H structure type. The crystallographic cell parameters are shown in Fig. 1. As Cu doping increases, the *c*-axis parameter increases systematically, as is the case for $Cu_xTiSe_2$ (25), indicating that the Cu is intercalated between the $TaS_2$ layers. Although impurity phases are not visible in any of the diffraction patterns up to x = 0.14, the fact that the growth of the *c*-axis stops near x = 0.12 indicates that this is the solubility limit for copper in the phase.

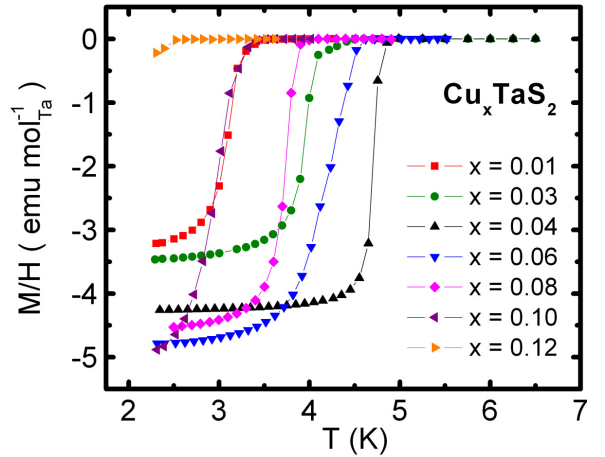

**Fig. 3** (Color online) Characterization of the superconducting transitions through measurement of the temperature dependent dc magnetizations in $Cu_xTaS_2$.

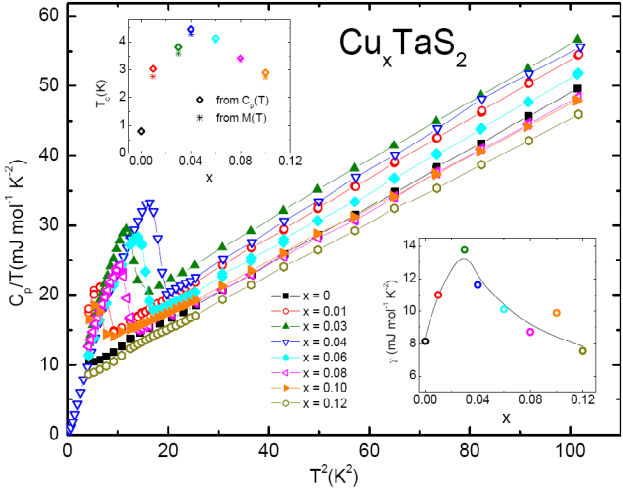
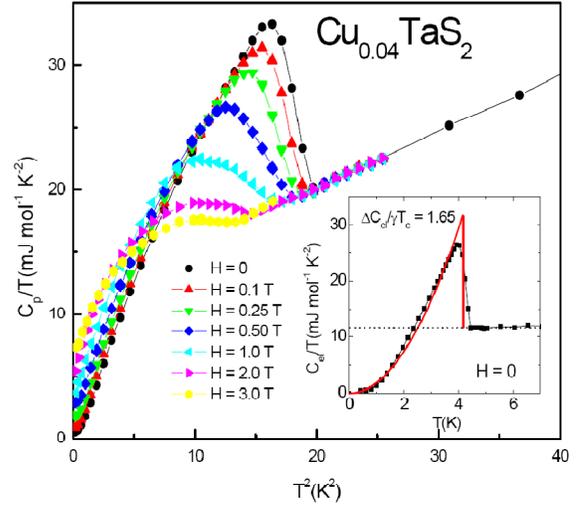

**Fig. 4** (Color online) Characterization of the superconducting transitions and electronic states in $Cu_xTaS_2$ for different x through measurement of the temperature-dependent specific heat. Lower inset: composition dependence of the electronic contribution to the specific heat in $Cu_xTaS_2$. Upper inset: $T_c$ values measured by specific heat and magnetization for $Cu_xTaS_2$, with the x = 0 value taken from the literature (18,19).

**Fig. 5** (Color online) Detailed characterization of the superconducting transition in the optimal superconducting composition $Cu_{0.04}TaS_2$ through specific heat. Inset: the H = 0 electronic contribution to the specific heat at the temperatures near the superconducting transition. The solid line shows the entropy conserving construction, with a fit to a $T^2$ polynomial at low temperatures.

The temperature dependent resistivities of polycrystalline pellets of $Cu_xTaS_2$ for the range of compositions synthesized are shown in Fig 2. All samples are metallic from 300 K to 2 K, and show a superconducting transition at low temperatures. The resistivity for undoped $TaS_2$ is also metallic over the whole temperature range, as has been reported previously (see e.g. ref 21). All samples show poor metallic behavior, with room temperature resistivities in the range of 1-4 mΩ-cm and residual resistivity ratios near 2. There is no systematic trend in the absolute magnitudes of the measured resistivities, which we attribute to the effects of grain boundaries and differing preferred orientations in the polycrystalline pellets. The samples show the presence of a superconducting transition in the resistivity at temperatures below 5 K; the inset to Fig. 2 shows the resistive transitions for x = 0.01, 0.04, and 0.08, indicating the increase and then decrease of $T_c$ with Cu doping.

The magnitudes of the low field dc magnetizations (Fig. 3) indicate the presence of bulk superconductivity in the phase pure samples. The observed transition temperatures are a systematic function of x: the $T_c$ first increases, to around 4.7 K for x = 0.04, and then decreases, until it is below 2 K for the 12 % Cu-doped sample. It is interesting to note that as little as 1 % doping pushes the $T_c$ of $2H$-$TaS_2$ up by about a factor of three - from less than 1 K to above 2.5 K, into a range of temperature where it is easily observed.

The low temperature specific heat characterization of the $Cu_xTaS_2$ system is shown in Fig. 4. The data show clear peaks at the superconducting transitions for different compositions, characteristic of bulk superconductivity. The $T_c$'s observed in these measurements are consistent with those observed in the susceptibility measurements of Fig. 3, and are summarized in the upper inset. Further, the specific heats at temperatures between $T_c$ and 10 K are well described as a sum of a $T^3$ phonon contribution and the T-linear electronic contribution, such that $C/T = \gamma + \beta T^2$. The β values are all very similar in the system, near 0.4 mJ mol$^{-1}$ K$^{-4}$, suggesting a Debye temperature of 165 K. The electronic contribution to the specific heat (lower inset, Fig. 4) first increases rapidly on doping, to a maximum of approximately 14 mJ mol$^{-1}$ K$^{-2}$ at x = 0.03, and then decreases on further doping to around 8 mJ mol$^{-1}$ K$^{-2}$ by the end of the chemical solid solution at 12 %. The highest value of γ is not found for the composition with the highest $T_c$.

The characterization of the superconducting transition by field-dependent specific heat and magnetization measurements for the optimal superconducting composition, $Cu_{0.04}TaS_2$, is shown in Fig. 5. The specific heat data plotted as

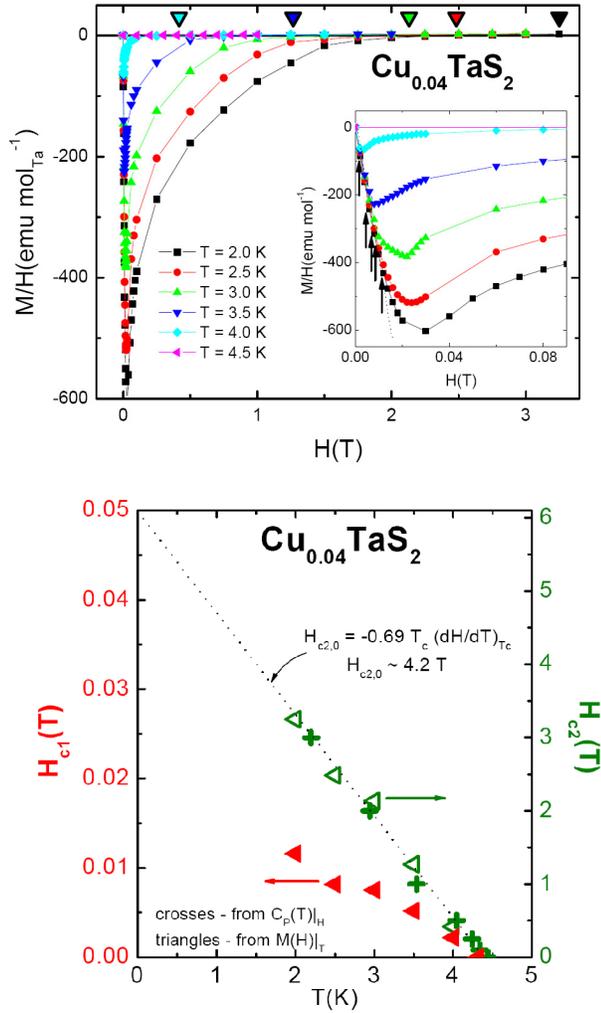
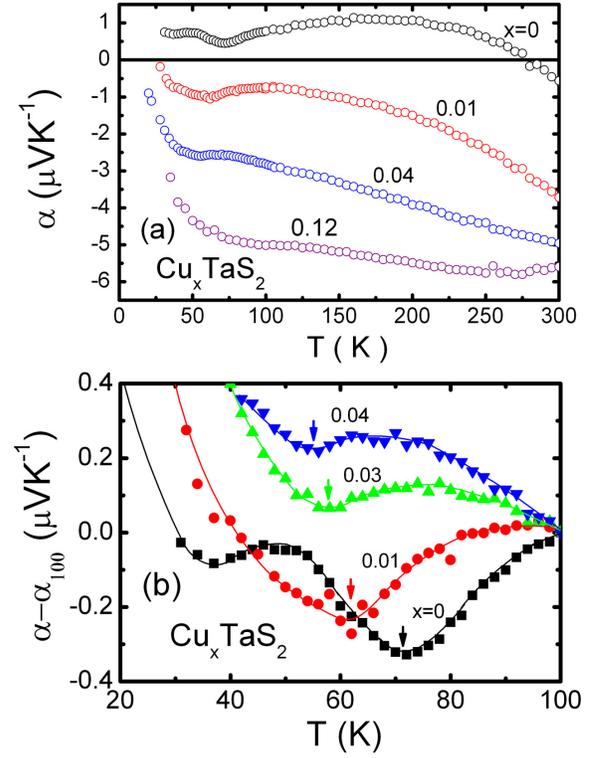

**Fig. 6** (Color online) Characterization of the superconducting state in $Cu_{0.04}TaS_2$. (a) M(H) isotherms, with the low-H part expanded in the inset. The estimated $H_{c2}$ and $H_{c1}$ values are indicated by large triangles and small vertical arrows respectively. (b) Estimated H – T phase diagram with $H_{c1}$ values (left axis) and $H_{c2}$ values (right axis) as determined from M(T,H) (full symbols) and $C_P(T,H)$ data (open symbols). The dotted line represents the linear fit close to $T_c$ according to the WWH formula.

$C_p/T$ vs. $T^2$ in Fig. 5 show the suppression of the transition in field, and indicate that $H_{c2}(0)$ is greater than 3 T. The inset shows the electronic specific heat $C_{el}$ at the superconducting transition with the phonon contribution subtracted. A BCS entropy-conserving construction (solid lines, inset Fig. 5) is used to determine the jump in the electronic specific heat at $T_c$. The fit shows that the sample is quite homogeneous, with a bulk $T_c$ of 4.2 K. Further it shows that the specific heat in the region of the superconducting transition is consistent with the form expected for a single gap s-wave BCS superconductor, and that $\Delta C_{el}/\gamma T_c = 1.65$, close to the ideal BCS ratio of 1.43.

**Fig. 7** (Color online) The temperature dependence of the Seebeck coefficients for $Cu_xTaS_2$. Upper panel, general behavior of the Seebeck coefficients over a wide temperature and x range. Lower panel, detail of the behavior for x=0, 0.01, 0.03, and 0.04 in the vicinity of the CDW transition. The lines are guides to the eye.

The M(H) data, shown in Fig. 6a, not only confirm the suppression of $T_c$ with applied magnetic field, but also allow us to estimate the upper and lower critical field values $H_{c2}$ and $H_{c1}$ as a function of temperature. The $H_{c2}$ values, where the sample enters the normal state at each temperature, are estimated as marked by large triangles. At low fields (inset, Fig. 6a), the M(H) isotherms are linear in H, as expected for a BCS type II superconductor; we estimate the $H_{c1}$ values, marked by vertical arrows, from the points where these curves deviate from linearity. Fig. 6b represents the H–T phase diagram for the optimal superconducting composition $Cu_{0.04}TaS_2$, with the full and open symbols representing values determined from the M(H) and the $C_p(T)$ data respectively. The Werthamer-Helfand-Hohenberg (WWH) equation (26) can be employed to determine $H_{c2}(0)$ from the H–T data close to $T_c$, i.e. $H_{c2}(0) \approx -0.69\ T_c\ (dH_{c2}/dT)|_{T_c}$, yielding an upper critical field value $H_{c2}(0) \approx 4.2$ T.

The CDW transition in 2H-TaS$_2$ has been observed previously in single crystals through resistivity measurements

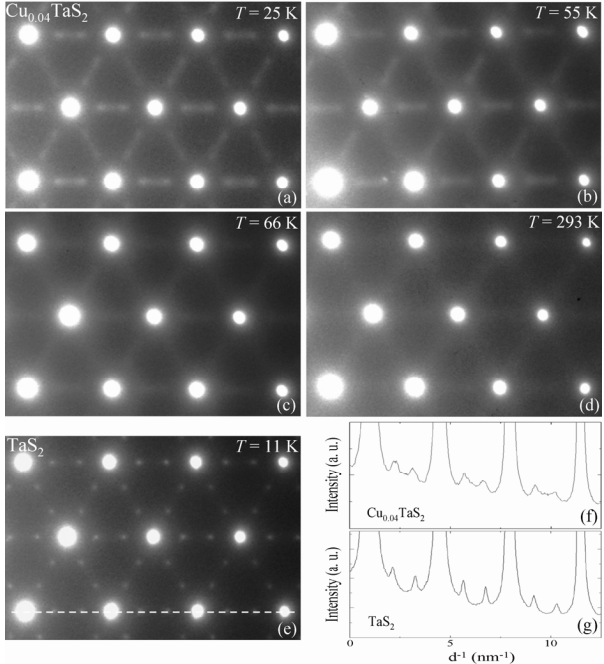

**Fig. 8** (a)-(d) Temperature evolution of the Electron Diffraction (ED) patterns in the hk0 plane for $Cu_{0.04}TaS_2$. Bright spots are the diffraction peaks from the hexagonal crystal structure. Diffuse scattering is present at high temperatures, but at 55 K and 25 K, the diffraction peaks from the CDW are clearly observed. (e) hk0 plane ED pattern at 11 K of $2H-TaS_2$. The weak, sharp CDW reflections are clearly shown between the fundamental reflections. (f) and (g) are the intensity profiles of the fundamental and the CDW reflections measured (along the dashed line in fig. 8e) from the ED patterns in figs 8a and 8e for $Cu_{0.04}TaS_2$ and $TaS_2$ respectively. Quantitative fits of the CDW peak intensity profiles yield a CDW q = (0.341 ± 0.003) a* for $TaS_2$, and a CDW q = (0.358 ± 0.004) a* for $Cu_{0.04}TaS_2$. The coherence length of the CDW is greater than 50 nm for $TaS_2$ and only ~ 4-5 nm for $Cu_{0.04}TaS_2$.

(21,27,28), and though it is observable through subtle changes in susceptibility and resistivity in our undoped polycrystalline $2H-TaS_2$ sample, we found that for the $Cu_xTaS_2$ system even 1 % doping made the transition more difficult to observe. It was below our detection limits in both resistivity and susceptibility measurements, but was observable through Seebeck coefficient measurements. These are shown in Fig. 7. The magnitudes of the Seebeck coefficients for all samples are small, ~1-6 μV K$^{-1}$, positive for undoped $TaS_2$ and negative for the Cu-doped samples. The data show (upper panel) that Cu doping systematically moves the Seebeck coefficient toward more negative values, suggesting that Cu acts at least in part as an n-type dopant. The lower panel shows the subtle changes observable at the CDW transition, which we can clearly follow up to a composition of x = 0.04, before they are no longer visible for x = 0.06. The data show that both the CDW and superconductivity are present in $Cu_xTaS_2$ at least up to the optimal superconducting composition of x = 0.04.

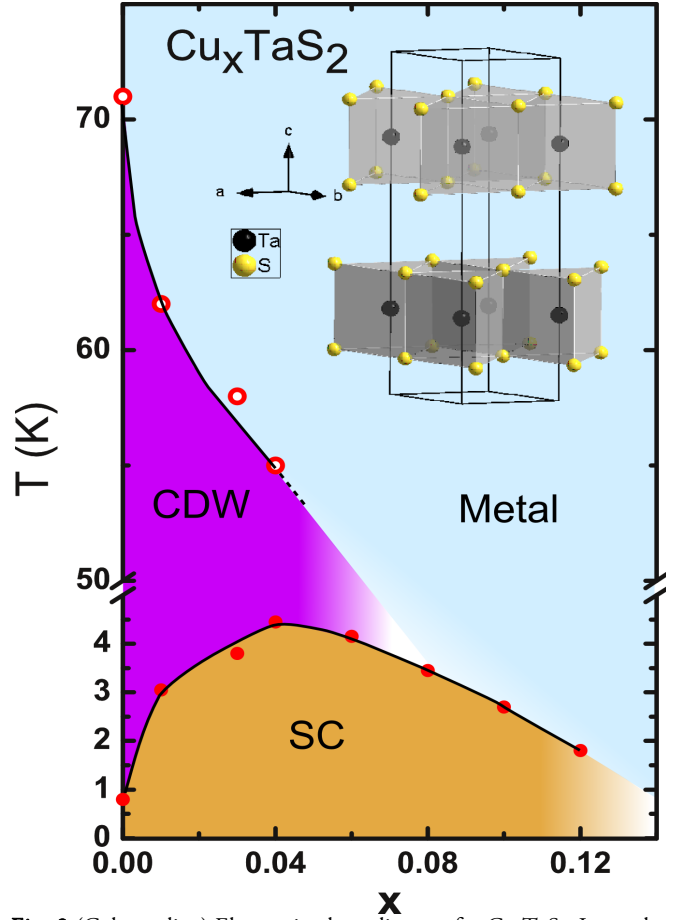

**Fig. 9** (Color online) Electronic phase diagram for $Cu_xTaS_2$. Inset: the $2H-TaS_2$ crystal structure. Copper is inserted between the $TaS_2$ planes.

We confirmed the presence of the CDW in the optimal superconducting compound $Cu_{0.04}TaS_2$ and the CDW transition temperature estimated from the Seebeck coefficient measurements by ED studies. The ED patterns in the (hk0) plane for $Cu_{0.04}TaS_2$, taken at different temperatures, are shown in Fig. 8(a)-8(d). The patterns show, *e.g.* at 293 K, the strong diffraction peaks from the hexagonal host structure. Some evidence of diffuse scattering between these peaks is present at high temperatures, likely related to the presence of a soft phonon, but between 66 K and 55 K weak and broad, but clear, peaks appear from the condensation of the CDW between those two temperatures.

For $Cu_xTiSe_2$ the q vector of the commensurate CDW, arising from a special relationship between valence band and conduction band states near the Fermi Energy (see, *e.g.* refs. 29-31), did not change with the Cu doping that induced the superconducting state for $Cu_xTiSe_2$, even though the CDW was destabilized (25). The CDW in $2H-TaS_2$ is more conventional in character, however, incommensurate and related to electron count and Fermi surface nesting (32). It is therefore possible that the q of the CDW will change in $Cu_xTaS_2$, as electrons are doped into the electronic system, through changing the size or shape of the Fermi surface. To test whether this is the case, we performed ED studies comparing undoped 2H-$TaS_2$ and the optimal superconducting composition, $Cu_{0.04}TaS_2$, at low temperatures. The CDW is well developed at both compositions. The ED patterns of the pure 2H-$TaS_2$ and the $Cu_{0.04}TaS_2$ samples are shown in Fig. 8(e) and Fig. 8(a), respectively. They reveal dramatic changes in the CDW on Cu doping. For $TaS_2$ itself (Fig. 8(e)), the CDW peaks are sharp, indicating a long CDW coherence length. Quantitative measurement of the diffracted intensities allows for fitting of the line profiles of the CDW reflections to determine the peak positions and the full widths at half maximum. The q of the CDW for $TaS_2$ is found to be (0.341(3),0,0), in good agreement with previous reports (q = 0.338). Further, the narrow peak width indicates that the coherence length for the CDW state is larger than 50 nm. For the optimally copper-doped superconducting sample (Fig. 8(a)), the CDW peaks are substantially more diffuse but still present. Quantitative fitting of the CDW peak profiles shows both a significant change in the q vector of the CDW, to (0.358(4), 0, 0), and a decrease in the coherence length of the CDW regions, by a factor of at least ten, to 4-5 nm. Thus the Cu doping has a clear, two-fold effect on the CDW. Firstly, the change in q indicates an electronic-doping-induced change in the Fermi surface compared to that of $TaS_2$ (though there may be a small additional contribution to the difference in q due to the different temperatures (11 K vs. 25 K) of the two diffraction patterns). Secondly, there is a dramatic decrease in the CDW coherence length, indicative of a disorder-induced disruption of the coherence of the CDW, which becomes limited to short distances on Cu doping. Our results on the characterization of the $Cu_xTaS_2$ system are summarized in the electronic phase diagram in Fig. 9.

## IV. DISCUSSION

The fact that $T_c$ is pushed above 2 K in 2H-$TaS_2$ by very low amounts of doping gives some insight into the many observations of superconductivity above 1 K in nominally pure 2H-$TaS_2$ and variants doped with organic molecules (23, 24). "$TaS_2$" samples with indications of $T_c$ above 1 K may in fact be very slightly nonstoichiometric through having small Ta excess or sub 1% quantities of impurity atoms present. It is not at all clear how an increase of $T_c$ by a factor of about three between $TaS_2$ and $Cu_{0.01}TaS_2$ can be possible if the effect of copper doping is only to change the carrier concentration by 1/100 of an electron per formula unit. There must be additional contributions - changing the interlayer coupling for example (33), or disrupting the ordering of the CDW state through impurity scattering due to the intercalated Cu, as seen in the change in CDW coherence length seen in electron diffraction. These factors may explain why $T_c$'s of 2.5 K are very commonly seen on intercalation of $TaS_2$ with neutral organic molecules. The current results suggest that the first increment in $T_c$ on doping 2H-$TaS_2$, up to 2.5 K from 0.8 K, comes from these initial effects, and that further increases, to the 4.5 K range, are due to the actual electronic doping of the system.

Despite overall similarities, there are significant differences between the $Cu_xTiSe_2$ and $Cu_xTaS_2$ systems. The sharp initial increase in the electronic contribution to the specific heat on Cu doping in $Cu_xTaS_2$ to a maximum at around x = 0.03, followed by a decrease at higher x, is in distinct contrast to what is observed in the $Cu_xTiSe_2$ system (25). In that case, γ increases continuously with Cu doping, and continues to increase for compositions beyond that of the optimal superconductor at $Cu_{0.08}TiSe_2$. Therefore the decrease in $T_c$ with increasing x cannot be due to a decreasing density of states. For $Cu_xTaS_2$ on the other hand, the substantial decrease in $T_c$ for Cu contents beyond the optimal 0.04 doping could be due to the 50 % decrease in the density of states for higher x. Despite similar $T_c$ values, the γ values at the optimal superconducting compositions are 12 mJ mol$^{-1}$K$^{-2}$ and 4 mJ mol$^{-1}$ K$^{-2}$ for $Cu_xTaS_2$ and $Cu_xTiSe_2$ respectively, and the zero K upper critical field of 4.2 T in $Cu_{0.04}TaS_2$ is approximately three times larger than that observed in $Cu_{0.08}TiSe_2$.

As for the copper oxide high-$T_c$ superconductors, where the competition between antiferromagnetism and superconductivity evolves as a function of electronic doping, the evolving balance between competing electronic states in CDW/superconducting systems is one of their most fundamentally interesting characteristics. This has been the subject of theoretical study for decades, and the analysis continues to evolve (see *e.g.* refs. 7-13). The details of exactly how CDWs and superconductivity compete at the microscopic level have not been experimentally established in the layered dichalcogenides to nearly the degree that the analogous competition has been established in the copper oxides, in part due to the unavailability of systems where the balance can be carefully titrated in chemically and structurally simple equilibrium compounds. Detailed experimental comparison between $Cu_xTiSe_2$ and $Cu_xTaS_2$, for example, where the chemistry, structure, and general superconductivity/CDW phase diagrams are very similar, but the Fermi surfaces and microscopic reasons for the presence of the CDW may be fundamentally different, could help to establish whether universal behavior exists for systems of this type.

ACKNOWLEDGEMENTS

The work at Princeton was supported by the US Department of Energy, Division of Basic Energy Sciences, grant DE-FG02-98ER45706. Work at Brookhaven National Laboratory was supported by the U.S. DOE/BES under Contract No. DE-AC02-98CH10886. EM acknowledges support from Rice University. TMM acknowledges the support of the NSF Graduate Research Fellowship Program.